# Metallic Bonding-Driven Elastic Softness and Optical Response in the Mg-Rich Laves-Phase $LaMg_2$

Farjana Mou[1,2], S. M. Nazmus Shakib Pias[3], M A Islam[4,5], Md Zahidur Rahaman[2,6,*]

[1]*Pabna Textile Engineering College, Bangladesh University of Textiles, Dhaka 1208*

[2]*EcoBengal Centre for Materials and Energy Research, Pabna 6600, Bangladesh*

[3]*Scientific Officer, Nuclear Safety Division, Bangladesh Atomic Energy Regulatory Authority (BAERA)*

[4]*Department of Physics, University of Barishal, Barishal 8200, Bangladesh*

[5]*Department of Electrical and Mechanical Engineering, Nagoya Institute of Technology, Gokiso-cho, Showa-ku, Nagoya, 466-8555, Japan*

[6]*School of Materials Science and Engineering, Faculty of Science, University of New South Wales, Sydney 2052, Australia*

[*]*Corresponding Author:*

*Md Zahidur Rahaman*

*Email: md_zahidur.rahaman@unswalumni.com | zahidur.rahaman@ecobengal.org*

**Abstract**

A systematic first-principles investigation of the structural, electronic, mechanical, and optical properties of the cubic C15 Laves-phase intermetallic compound $LaMg_2$ is performed within density functional theory. The calculated elastic constants satisfy the mechanical stability criteria for cubic crystals, confirming the intrinsic stability of the C15 phase. $LaMg_2$ exhibits relatively low bulk, shear, and Young's moduli, indicating enhanced compressibility and elastic softness compared with transition-metal-based Laves phases. Direction-dependent elastic analysis reveals moderate anisotropy in Young's modulus, shear modulus, and Poisson's ratio, whereas linear compressibility remains nearly isotropic, consistent with the high crystallographic symmetry. The ductile nature of $LaMg_2$ is supported by Pugh's ratio and Poisson’s ratio, suggesting resistance to brittle failure and dominance of metallic bonding. Electronic structure calculations confirm metallic behavior with finite density of states at the Fermi level, primarily originating from La-5d states, accompanied by delocalized charge density characteristic of metallic interactions. The optical response further reflects the metallic nature through high reflectivity, strong optical conductivity at low photon energies, and pronounced absorption in the ultraviolet region. The combination of mechanical compliance, ductility, and metallic optical response highlights $LaMg_2$ as a promising lightweight intermetallic material for applications requiring structural stability, damage tolerance, and efficient electromagnetic shielding or reflective components.

## 1. Introduction

Magnesium (Mg) has emerged as a significant element in engineering due to its abundant presence on Earth. Its widespread availability, coupled with its low density (around 1.74 gm/cm$^3$), remarkable specific strength, and excellent rigidity compared to various other engineering materials like aluminium, steel, and polymer-based composites, has garnered substantial interest in magnesium alloys [1, 2]. Beyond these qualities, magnesium possesses other favourable characteristics, such as effective electromagnetic shielding, high damping capacity, strong thermal conductivity, ease of machining, and a high potential for recycling [3]. These alloys have found diverse applications in the transportation sector, notably within the automotive industry and aerospace manufacturing, owing to their appealing attributes. This has led to magnesium being recognized as an environmentally friendly green engineering material. In the past few years, there has been significant interest in magnesium alloys because of their possible utilization as materials for storing hydrogen and as the lightest option for constructing metal structures. Exploring the phase relationships of these alloys, both in conditions of equilibrium and non-equilibrium, is highly crucial for designing novel alloy compositions and improving manufacturing methods. These appealing traits of magnesium have served as a catalyst for our present investigation on Mg-based alloy.

A set of seventeen silvery-white lustrous soft heavy metals are known as rare-earth elements. Materials composed of rare earth elements have a variety of applications in practice from electrical/electronic components to geochemistry and various industrial processes. In the earth's crust rare earth elements are relatively abundant in spite of their name. Most of the rare earths are used as catalysts and magnets [4]. These are also used in polishing, glass, batteries, ceramics, metallurgy, and phosphors [5]. Among rare-earth elements lanthanum (La) is one of the most plentiful elements (39 ppm) in earth's crust resulting in a broad application in various sector, such as hydrogen storage, alkali-resistant glass, battery-electrodes, lenses, camera, and oil refineries [6]. La and Ce are also used as catalysts and diesel additives and in petroleum refining [6]. La also has important applications in medical science. Lanthanum carbonate ($La_2(CO_3)_3$) is utilized as a phosphate binder in cases of high levels of phosphate in the blood seen in kidney failure patients.

The Laves phase group is one of the most studied oldest family of compounds with a diverse range of physical, chemical, magnetic, and superconducting characteristics [7]. These intermetallic compounds have garnered significant interest due to their intriguing characteristics, which encompass polymorphism as well as distinct magnetic and electrical properties [8]. These intermetallic compounds exhibit excellent resistance to corrosion and creep and possess functional properties in the realm of magnetism and electricity, enabling their use as magnetic substances, magneto-optical elements, and materials for storing hydrogen [9, 10].

The intriguing properties La, Mg, and Laves phase family as discussed above motivate us to investigate the detailed physical properties of $LaMg_2$ Laves phase intermetallic compound. Herein theoretical investigations have been conducted on $LaMg_2$ Laves phase compound and a thorough comparison has been made of the observed properties with other Laves phase compounds to check the feasibility of the studied one.

**2. Computational Methodology**

First-principles calculations based on density functional theory (DFT) were performed using the plane-wave pseudopotential formalism as implemented in the CASTEP package [11]. The exchange-correlation interaction was treated within the generalized gradient approximation (GGA) employing the Perdew-Burke-Ernzerhof (PBE) functional [12-14]. The electronic configurations considered in the pseudopotential generation were La $5s^2$ $5p^6$ $5d^1$ $6s^2$ and Mg $2s^2$ $2p^6$ $3s^2$. A plane-wave kinetic energy cutoff of 500 eV was adopted to ensure reliable convergence of the total energy. Brillouin zone integrations were conducted using the Monkhorst-Pack scheme with a $12 \times 12 \times 8$ k-point mesh (312 irreducible k-points) for the primitive unit cell of $LaMg_2$. Structural optimizations were performed using the Broyden-Fletcher-Goldfarb-Shanno (BFGS) algorithm until the residual forces and stresses satisfied strict convergence criteria [15]. Specifically, the thresholds were set to $1.0 \times 10^{-5}$ eV/atom for total energy, 0.03 eV/Å for atomic forces, 0.05 GPa for stress, and 0.001 Å for atomic displacements. The elastic stiffness constants of the cubic Laves-phase $LaMg_2$ were evaluated using the stress-strain approach. In these calculations, tighter convergence tolerances were employed, with limits of $2.0 \times 10^{-6}$ eV/atom for total energy, $2.0 \times 10^{-4}$ Å for maximum ionic displacement, and $6.0 \times 10^{-3}$ eV/Å for maximum force. A maximum strain amplitude of 0.003 was applied. Slightly different convergence criteria were used for geometry optimization and elastic constant calculations to ensure optimal numerical accuracy for each computational task.

This strategy enabled highly reliable determination of both equilibrium structural parameters and elastic properties.

## 3. Results and Discussion

### 3.1 Structural Properties

**Figure 1** illustrates the crystal structure of the rare-earth Laves-phase intermetallic $LaMg_2$, shown in both the conventional and primitive unit-cell representations. $LaMg_2$ crystallizes in the cubic C15 Laves-phase structure, which is characteristic of $AB_2$-type intermetallic compounds. This structure belongs to the face-centered cubic crystal system and is described by the space group Fd-3m (No. 227) [16]. In the conventional unit cell (**Figure 1(a)**), La atoms occupy the A-sites of the Laves phase and form a diamond-like sublattice, while Mg atoms populate the B-sites, constructing a three-dimensional network of corner-sharing Mg tetrahedra [17]. The coordination environment reflects the typical topology of the C15 structure. This highly symmetric atomic arrangement underpins the structural stability of $LaMg_2$. The primitive unit cell (**Figure 1(b)**) highlights the minimal repeating structural motif required to reproduce the full crystal lattice through translational symmetry. The reduced cell preserves the same atomic connectivity and symmetry elements as the conventional cell, while containing fewer atoms, making it particularly suitable for first-principles calculations. Owing to its cubic symmetry, $LaMg_2$ is characterized by a single independent lattice parameter ($a$ = 8.819012 Å), with all interaxial angles equal to 90º. The unit-cell volume is therefore given by $V = a^3$ (685.8984 Å$^3$). This calculated structural parameters agree well with the experimental parameters implying the reliability of the present calculations [18].

### 3.2 Electronic Properties

**Figure 2** presents the electronic structure of $LaMg_2$ through its Brillouin zone, band dispersion, partial density of states (PDOS), and total density of states (TDOS), providing a comprehensive description of its electronic behavior. **Figure 2(a)** shows the first Brillouin zone corresponding to the cubic C15 Laves-phase structure, together with the high-symmetry points and directions used for band-structure calculations. The selected path connects the standard symmetry points W-L-Γ-X-W-K, ensuring a representative sampling of the electronic states along principal crystallographic directions. The calculated electronic band structure of $LaMg_2$ (**Figure 2(b)**) reveals multiple bands crossing the Fermi level ($E_F$) unequivocally indicate the metallic nature of $LaMg_2$. The dispersive character of several bands near the Fermi level suggests the presence of mobile charge carriers, consistent with good electrical conductivity. Notably, the band dispersion is relatively smooth along different symmetry directions, reflecting the high structural symmetry of the cubic lattice.

Further insight into the orbital contributions is provided by the PDOS shown in **Figure 2(c)**. The electronic states near the Fermi level are dominated primarily by La-5d states, with additional contributions from Mg-2p orbitals. In contrast, the La-6s and Mg-3s states contribute mainly at energies further away from $E_F$, while La-5p states are located deeper in the valence band. The substantial overlap between La-5d and Mg-derived states indicates significant hybridization, pointing to mixed metallic bonding character between La and Mg atoms. The TDOS of $LaMg_2$ (**Figure 2(d)**) further confirms its metallic behavior, as a finite density of states persists at the Fermi level. The relatively high TDOS (6.90 states/eV) around $E_F$ suggests a robust availability of electronic states for conduction, which is characteristic of metallic Laves-phase intermetallics. Additionally, the broad distribution of states across the valence and conduction regions reflects the delocalized nature of the electronic structure.

**Figure 3** presents the electronic charge distribution of $LaMg_2$, providing direct insight into its bonding characteristics and charge localization within the C15 Laves-phase lattice. **Figure 3(a)** displays the total electron density isosurface within the conventional cubic unit cell. The charge density is predominantly localized around the atomic cores, with more pronounced accumulation surrounding the La atoms compared to Mg. The isosurface topology reveals continuous electron density between neighbouring atoms, indicating the presence of delocalized metallic bonding rather than strongly directional covalent interactions. The spatial continuity of the charge distribution throughout the lattice further supports the metallic nature inferred from the electronic band structure. **Figure 3(b)** illustrates a planar slice through the unit cell highlighting the relative position of La atoms and the corresponding charge-density distribution. Regions of enhanced electron density (blue contours) are centered primarily on La sites. The absence of strong directional charge accumulation along specific La-La bonds suggests that bonding is governed predominantly by metallic interactions with limited covalent character. The relatively symmetric charge distribution around La reflects the high cubic symmetry of the C15 framework. **Figure 3(c)** shows a contour map taken along an inclined crystallographic plane, enabling visualization of charge-density variations across both La and Mg sublattices. The charge density contour reveals negligible charge overlap between neighbouring La and Mg atoms, consistent with the weak orbital hybridization identified in the PDOS analysis. However, the charge density remains largely centered on atomic sites, indicating that electron sharing is delocalized rather than localized in directional bonds. **Figure 3(d)** presents a two-dimensional contour map of the total electron density. The nearly isotropic distribution of charge density across the plane highlights the homogeneous electronic environment characteristic of cubic intermetallic systems. The smooth and continuous electron-density background between atomic sites reflects the presence of mobile electrons contributing to metallic conductivity. Moreover, the absence of pronounced charge depletion or accumulation regions between specific atom pairs further confirms that $LaMg_2$ exhibits predominantly metallic bonding with minor ionicity arising from electronegativity differences between La and Mg.

Overall, the electronic structure analysis demonstrates that $LaMg_2$ is a metallic intermetallic compound (delocalized nature) with electronic states at the Fermi level primarily governed by La-5d orbitals, supplemented by Mg-2p contributions. This bonding topology is intrinsic to the C15 Laves-phase architecture and plays a decisive role in stabilizing the structure while governing its mechanical and transport properties.

### 3.3 Optical Properties

**Figure 4** summarizes the frequency-dependent optical response of $LaMg_2$ over a broad photon-energy range, providing insight into its interband transitions, free-carrier behavior, and collective electronic excitations. The reflectivity spectrum (**Figure 4(a)**) exhibits a relatively high value in the low-energy region, consistent with the metallic nature of $LaMg_2$. The pronounced reflectance at small photon energies arises from intraband (Drude-like) free-carrier contributions associated with the partially filled bands crossing the Fermi level. As the photon energy increases, the reflectivity decreases sharply, indicating reduced free-carrier screening and the onset of significant interband transitions. At higher energies, weak oscillatory features emerge, corresponding to electronic excitations between deeper valence states and unoccupied conduction bands. The absorption coefficient (**Figure 4(b)**) shows substantial intensity across a wide energy range, reaching magnitudes on the order of $10^5$ $cm^{-1}$. The absence of a well-defined absorption edge further confirms the metallic character, as optical transitions can occur at arbitrarily low photon energies due to the lack of a bandgap. Distinct absorption peaks in the intermediate and high-energy regions originate from interband transitions, primarily involving La-5d states near the Fermi level and deeper-lying Mg- and La-derived states, consistent with the electronic structure analysis. The complex refractive index (**Figure 4(c)**) reflects strong dispersion at low photon energies. The real part of the refractive index, $n(\omega)$, exhibits a high static value that rapidly decreases with increasing photon energy, characteristic of metallic systems with strong free-electron response. The imaginary component, $k(\omega)$, follows a similar decay trend, indicating substantial optical attenuation in the low-energy regime due to free-carrier absorption. Further insight is obtained from the complex dielectric function $\varepsilon(\omega) = \varepsilon_1(\omega) + i\varepsilon_2(\omega)$ (**Figure 4(d)**). As energy increases, $\varepsilon_1(\omega)$ decreases and approaches zero, marking the screened plasma frequency where the material transitions from reflective to transmissive behavior. The imaginary part, $\varepsilon_2(\omega)$, displays pronounced peaks corresponding to interband electronic excitations, in agreement with features observed in the absorption spectrum. The optical conductivity spectrum (**Figure 4(e)**) reinforces this interpretation. The real part of the conductivity exhibits a strong low-energy response attributable to free carriers, followed by additional structures at higher photon energies arising from interband transitions. The imaginary part shows dispersive features associated with polarization processes and collective carrier dynamics. Finally, the energy-loss function (**Figure 4(f)**) reveals a prominent peak corresponding to the bulk plasmon resonance. This peak defines the characteristic plasma frequency of $LaMg_2$ and signifies the collective oscillation of conduction electrons. The position of this maximum is consistent with the zero-crossing of $\varepsilon_1(\omega)$, confirming the metallic plasmonic response of the compound.

Overall, the optical spectra demonstrate that $LaMg_2$ exhibits a pronounced free-electron response at low photon energies combined with significant interband activity at higher energies. The strong reflectivity, high optical conductivity, and well-defined plasmon resonance collectively corroborate its metallic electronic structure. These characteristics suggest that, beyond its structural and mechanical robustness, $LaMg_2$ may exhibit potential for applications requiring metallic reflectors, plasmonic materials, or high-conductivity intermetallic components operating in the ultraviolet to visible energy range.

### 3.4 Mechanical Properties

The calculated elastic constants and derived mechanical parameters of $LaMg_2$ are summarized in **Table 1** and compared with other representative Laves-phase and related intermetallic compounds. For a cubic crystal, the mechanical stability criteria require that $C_{11} > 0$, $C_{44} > 0$, and $C_{11} - C_{12} > 0$. The obtained elastic constants for $LaMg_2$ ($C_{11}$ = 55.50 GPa, $C_{12}$= 25.94 GPa, and $C_{44}$ = 23.24 GPa) satisfy these conditions, confirming its mechanical stability within the C15 framework. The moderate magnitude of $C_{11}$ reflects the resistance of the lattice to longitudinal compression along the principal crystallographic axes, while the comparatively smaller $C_{44}$ indicates a lower resistance to shear deformation. The small value of $C_{11}$- $C_{12}$ further suggests limited elastic anisotropy in the cubic phase, consistent with the isotropic bonding environment inferred from the electronic and charge-density analyses. The value of $C_{11}$- $C_{12}$ results in a Zener anisotropy factor (*A* = 1.57) that suggests noticeable elastic anisotropy in the cubic phase, indicating that mechanical responses vary moderately across different crystallographic directions [19].

The calculated bulk modulus *B* (35.80 GPa) indicates that $LaMg_2$ possesses relatively high compressibility compared with transition-metal-based Laves phases such as $MgCu_2$ or $Rh_2La$, which exhibit significantly larger bulk moduli. This reduced incompressibility can be attributed to the weaker metallic bonding strength associated with La-Mg interactions compared to systems containing more strongly bonding d-electron elements. The shear modulus *G* (19.38 GPa) and Young's modulus *E* (49.26 GPa) further confirm the comparatively soft mechanical character of $LaMg_2$. In contrast to stiffer intermetallics such as $Be_2Nb$ or $MgNi_2$, $LaMg_2$ exhibits lower resistance to both shear and uniaxial deformation. This mechanical softness is consistent with its delocalized metallic bonding and moderate orbital hybridization, as revealed by the electronic structure analysis. The ductility of $LaMg_2$ can be assessed using Pugh's ratio *B*/*G*, which is calculated to be 1.84. Since this value exceeds the critical threshold of ~1.75, $LaMg_2$ is predicted to exhibit ductile behavior [20]. This conclusion is further supported by the Poisson's ratio ($\nu = 0.27$), which lies within the typical range for metallic systems and indicates a significant contribution from central (metallic) interatomic forces rather than strong directional covalent bonding [15, 21].

In general, $LaMg_2$ can be classified as a mechanically stable, relatively soft, and moderately ductile intermetallic compound. Compared to other members of the Laves-phase family, it demonstrates lower stiffness and incompressibility but retains sufficient mechanical robustness for structural stability. The combination of metallic bonding, moderate elastic moduli, and predicted ductility may be advantageous in applications where mechanical compliance and resistance to brittle fracture are desirable, particularly in conjunction with its metallic electronic and optical characteristics.

**3.5 Comparative Elastic Behavior**

**Figure 5** presents a comparative overview of the elastic properties, bulk modulus (B), shear modulus (G), and Young's modulus (E), for a selection of Laves-phase intermetallic compounds. These three moduli collectively describe the resistance of a material to volumetric compression (B), shear deformation (G), and uniaxial elastic deformation (E), and thus provide a comprehensive measure of mechanical rigidity and structural robustness. Within this comparison, $LaMg_2$ exhibits moderate values of all three elastic moduli relative to the other Laves phases included in **Figure 5**. The bulk modulus of $LaMg_2$ is noticeably lower than that of transition-metal-rich compounds such as $Be_2Nb$, $Rh_2La$, and $MgNi_2$, indicating a comparatively higher compressibility. This behavior can be attributed to the dominance of Mg-Mg and La-Mg bonding interactions, which are generally less stiff than the more covalent or d-electron-mediated bonding present in compounds containing late transition metals. Similarly, the shear modulus of $LaMg_2$ remains relatively small, reflecting a limited resistance to shape change under applied shear stress. This suggests that $LaMg_2$ is mechanically softer and more prone to shear deformation compared to high modulus Laves phases, particularly those containing Rh or Ni. The corresponding Young's modulus follows the same trend, further confirming the overall elastic softness of $LaMg_2$.

Despite not ranking among the stiffest Laves-phase compounds, the mechanical response of $LaMg_2$ is internally consistent and characteristic of Mg-based intermetallics, where low density and elastic compliance are often favoured over extreme stiffness. The combination of modest bulk, shear, and Young's moduli implies that $LaMg_2$ may offer advantages in applications where mechanical flexibility, reduced elastic mismatch, or lightweight structural components are desirable, rather than maximum hardness or rigidity.

### 3.6 Directional Elastic Anisotropy

**Figure 6** illustrates the directional dependence of the elastic response of $LaMg_2$ through two-dimensional (2D) and three-dimensional (3D) representations of Young's modulus, linear compressibility, shear modulus, and Poisson's ratio, while the corresponding extrema and anisotropy factors are summarized in **Table 2**. Despite its cubic C15 symmetry, $LaMg_2$ exhibits moderate elastic anisotropy, reflecting direction-dependent deformation behavior. As shown in **Figure 6(a)**, Young's modulus varies between 38.98 and 57.31 GPa, yielding an anisotropy ratio 1.47. This variation indicates that the stiffness of $LaMg_2$ depends on crystallographic orientation, consistent with the moderate magnitude of Young's modulus and the relatively low resistance to uniaxial deformation discussed earlier. The deviation from spherical symmetry in the 3D surface highlights preferred directions of mechanical compliance within the lattice. In contrast, the linear compressibility displayed in **Figure 6(b)** is nearly isotropic, as evidenced by identical minimum and maximum values of 9.31 $TPa^{-1}$ and an anisotropy factor 1 (**Table 2**). This isotropic volumetric response is characteristic of cubic systems and supports the conclusion that $LaMg_2$ maintains mechanical stability under hydrostatic pressure despite its relatively high compressibility compared with transition-metal-rich Laves phases. The shear modulus distributions shown in **Figure 6(c)** reveal more pronounced anisotropy, with values ranging from 14.78 to 23.24 GPa and an anisotropy ratio 1.57. This directional dependence of shear resistance is consistent with the relatively low shear modulus, indicating that $LaMg_2$ is more susceptible to shear deformation along specific crystallographic planes. Poisson's ratio exhibits the strongest anisotropy among the elastic parameters (**Figure 6(d)**), spanning a wide range from 0.103 to 0.419 and yielding a high anisotropy factor 4.05. Such behavior reflects a highly direction-dependent transverse strain response and underscores the mixed metallic bonding nature of $LaMg_2$. The presence of relatively high Poisson's ratio values along certain directions further supports the ductile character predicted from the Pugh's ratio and the average Poisson's ratio value of 0.27.

In general, the combined analysis of **Figure 6** and **Table 2** confirms that $LaMg_2$, while elastically soft on average, exhibits moderate anisotropy in its directional mechanical response, particularly in shear deformation and transverse strain. These findings are fully consistent with the previously discussed elastic constants, bonding characteristics, and predicted ductility, reinforcing the classification of $LaMg_2$ as a mechanically stable and moderately ductile Laves-phase intermetallic compound.

## 4. Conclusions

In summary, a comprehensive density functional theory study has been carried out to elucidate the structural, electronic, mechanical, and optical properties of the rare-earth C15 Laves-phase intermetallic $LaMg_2$. The compound is confirmed to be mechanically stable within the cubic Fd-3m framework and exhibits relatively low bulk, shear, and Young's moduli, indicating elastic softness and enhanced compressibility compared with transition-metal-rich Laves phases. Directional elastic analysis reveals moderate anisotropy in Young's modulus, shear modulus, and Poisson's ratio, while linear compressibility remains nearly isotropic, consistent with the high crystallographic symmetry. The predicted ductile behavior, supported by Pugh's ratio and Poisson's ratio, reflects the dominance of delocalized metallic bonding, as corroborated by electronic band structure and charge-density analyses showing metallic character governed primarily by La-5d states. The optical response further confirms this metallic nature through high reflectivity, strong low-energy optical conductivity, and a well-defined plasmonic feature. Collectively, these results establish $LaMg_2$ as a mechanically compliant, moderately ductile, and electronically metallic intermetallic compound, highlighting its potential for lightweight structural components and functional applications requiring metallic conductivity, optical reflectivity, and electromagnetic shielding.

**CRediT Authorship Contribution Statement**

**Farjana Mou:** Conceptualization, Methodology, Investigation, Software, Simulation, Writing - original draft. **S. M. Nazmus Shakib Pias:** Writing - review & editing, Validation. **M A Islam:** Writing - review & editing, Validation. **Md Zahidur Rahaman:** Conceptualization, Methodology, Simulation, Writing-review and editing, Supervision, Validation.

**Declaration of Competing Interest**

The authors declare that they have no known competing financial interests or personal relationships that could have appeared to influence the work reported in this paper.

**Data Availability**

The authors declare that the data supporting the findings of this study are provided within the paper. Additional raw data files in alternative formats are available from the corresponding author upon reasonable request.

**Figures**

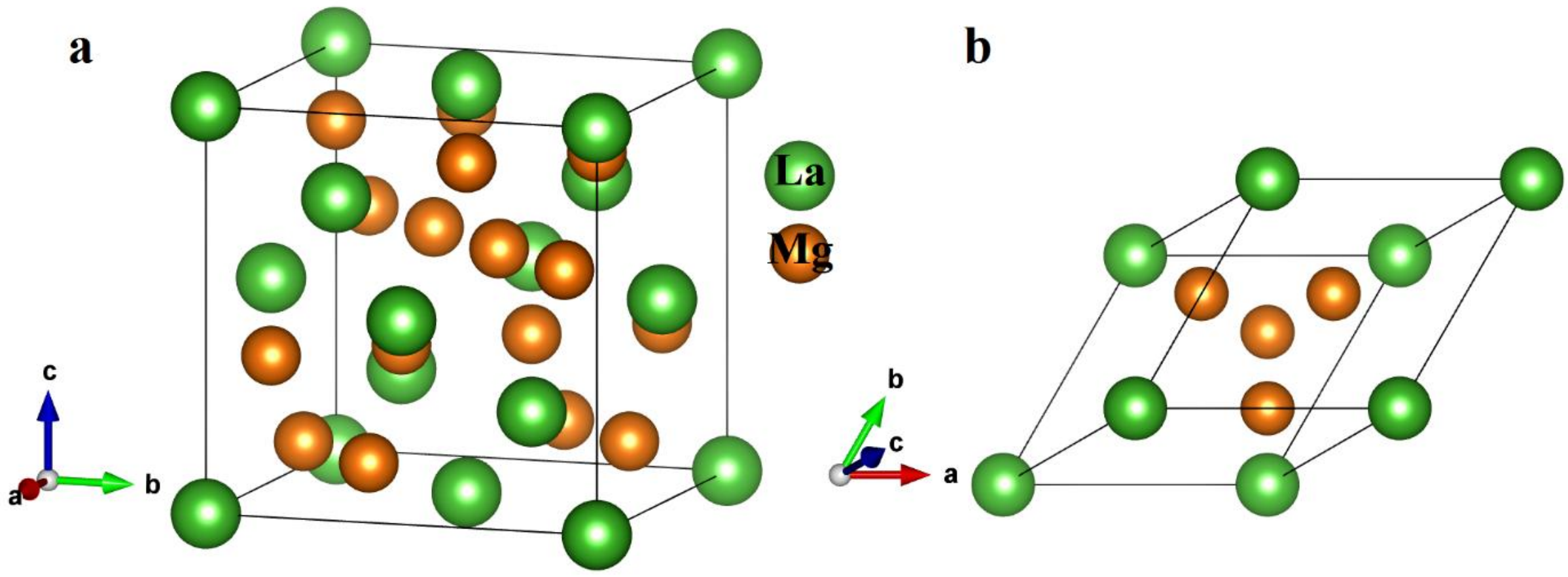

**Figure 1:** (a) Crystal structure (conventional unit cell) of rare-earth Laves phase $LaMg_2$. (b) Primitive unit cell of $LaMg_2$.

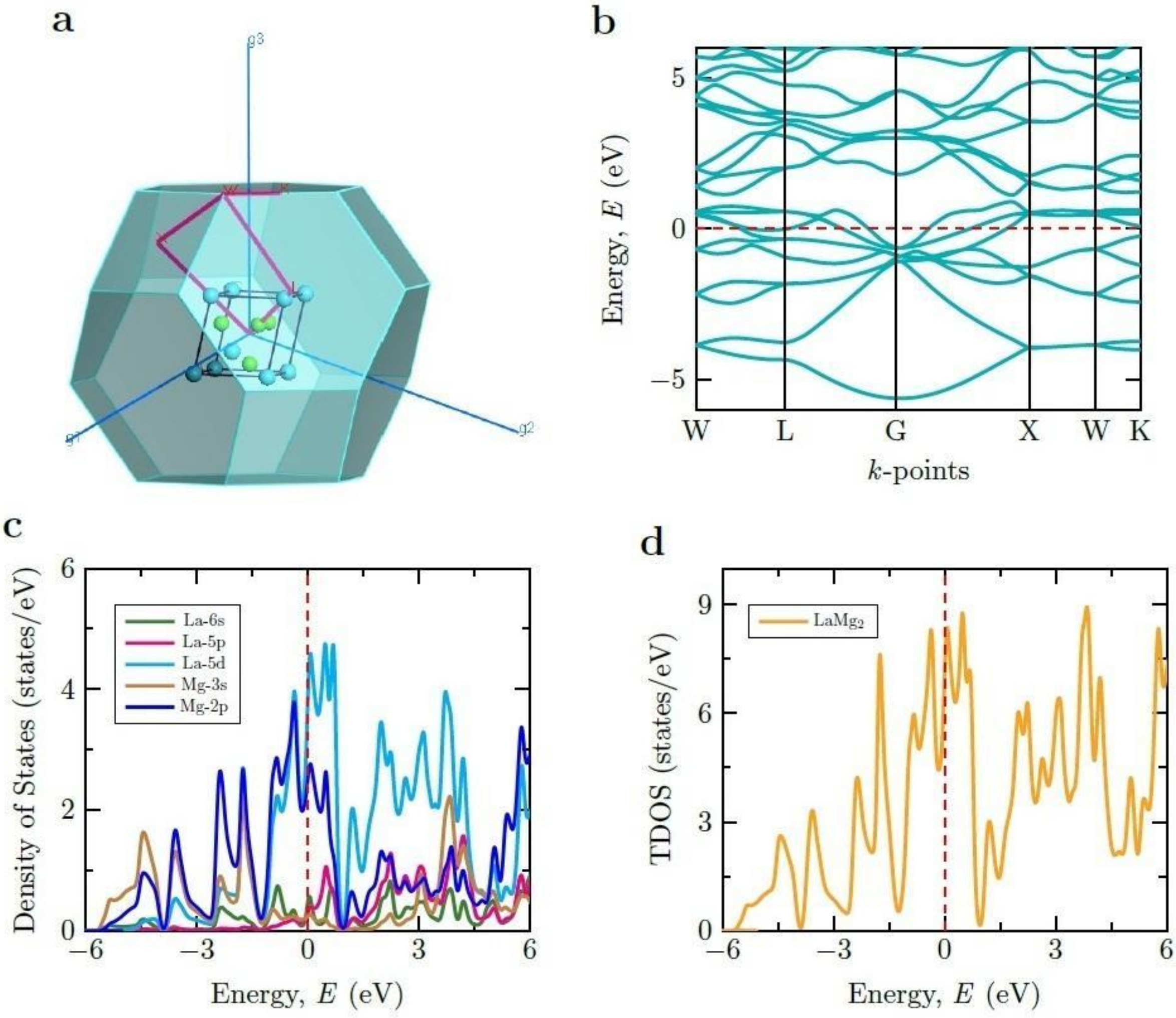

**Figure 2:** (a) First Brillouin zone of $LaMg_2$ showing high symmetry labels for high symmetry lines and points. (b) Electronic band structure of $LaMg_2$. (c) Partial density of states (PDOS) of $LaMg_2$. (d) Total density of states (TDOS) of $LaMg_2$.

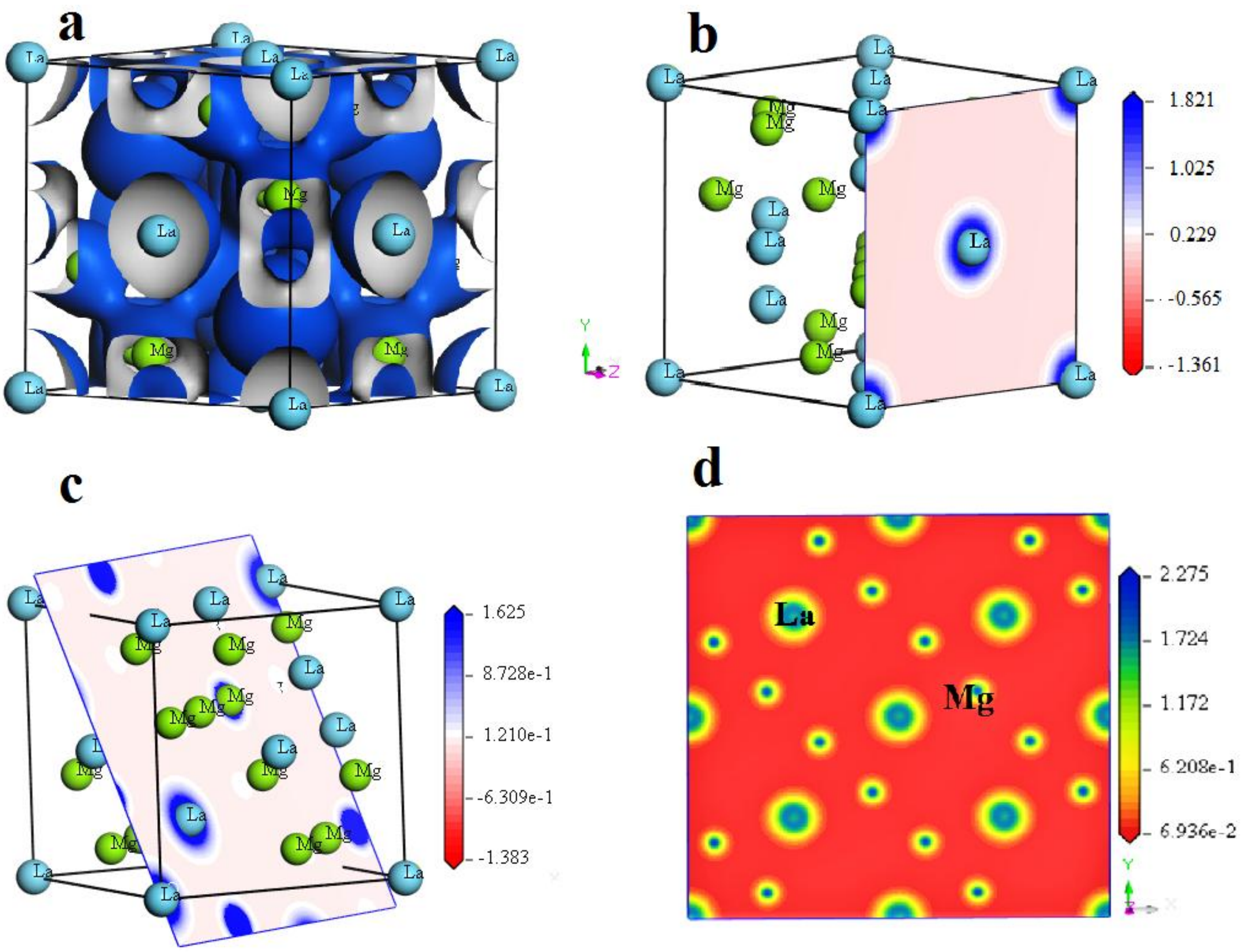

**Figure 3:** (a) The total electron density isosurface of $LaMg_2$. (b) The relative position of La in the crystal structure. (c) The relative position of contour map showing charge density of La and Mg atom. (d) Contour map of total electron density for $LaMg_2$.

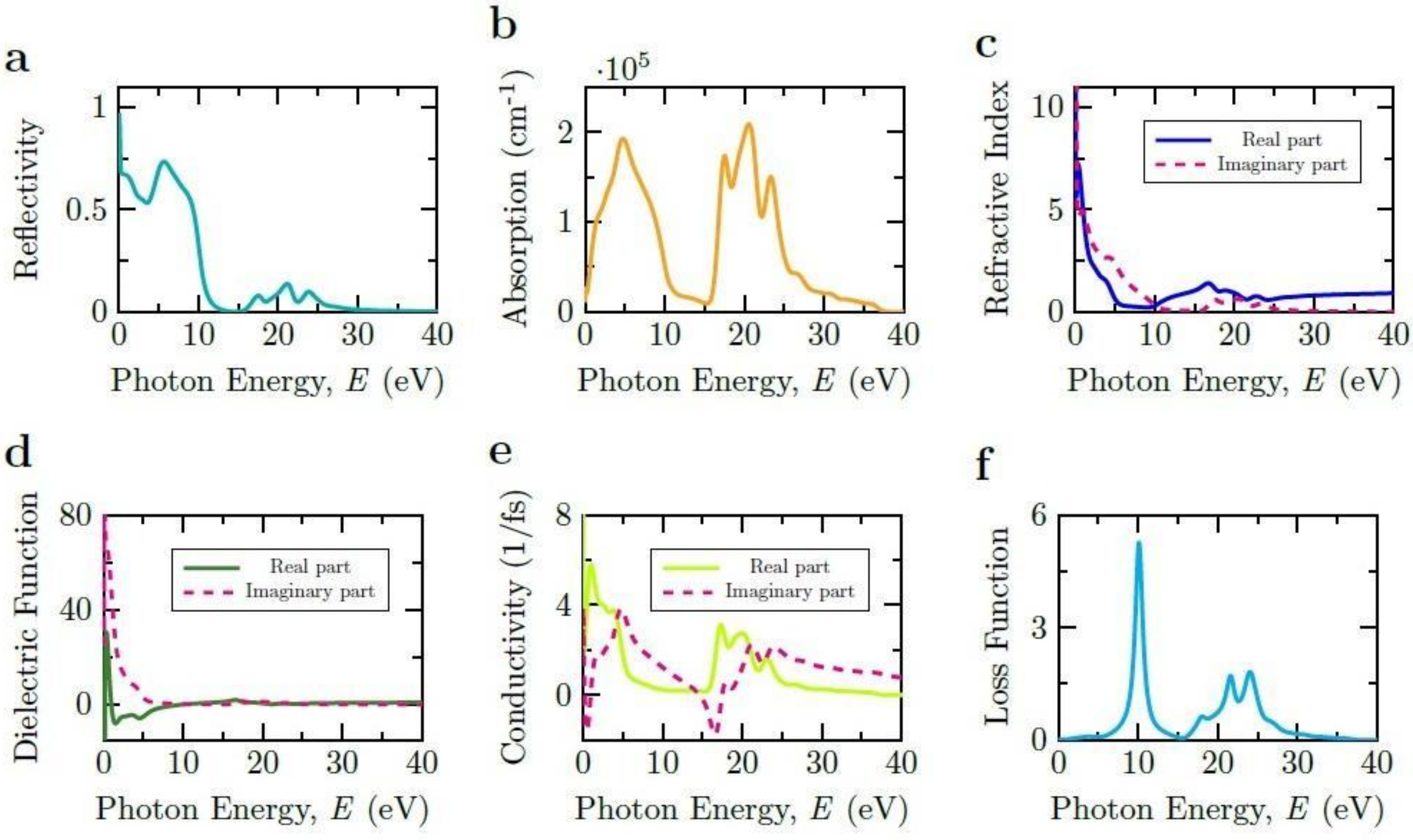

**Figure 4:** Optical proprieties of $LaMg_2$ Laves phase. (a) Reflectivity, (b) Absorption coefficient, (c) Refractive index, (d) Real and imaginary part of the dielectric function, (e) Conductivity, and (f) Loss function, of $LaMg_2$ rare-earth Laves phase.

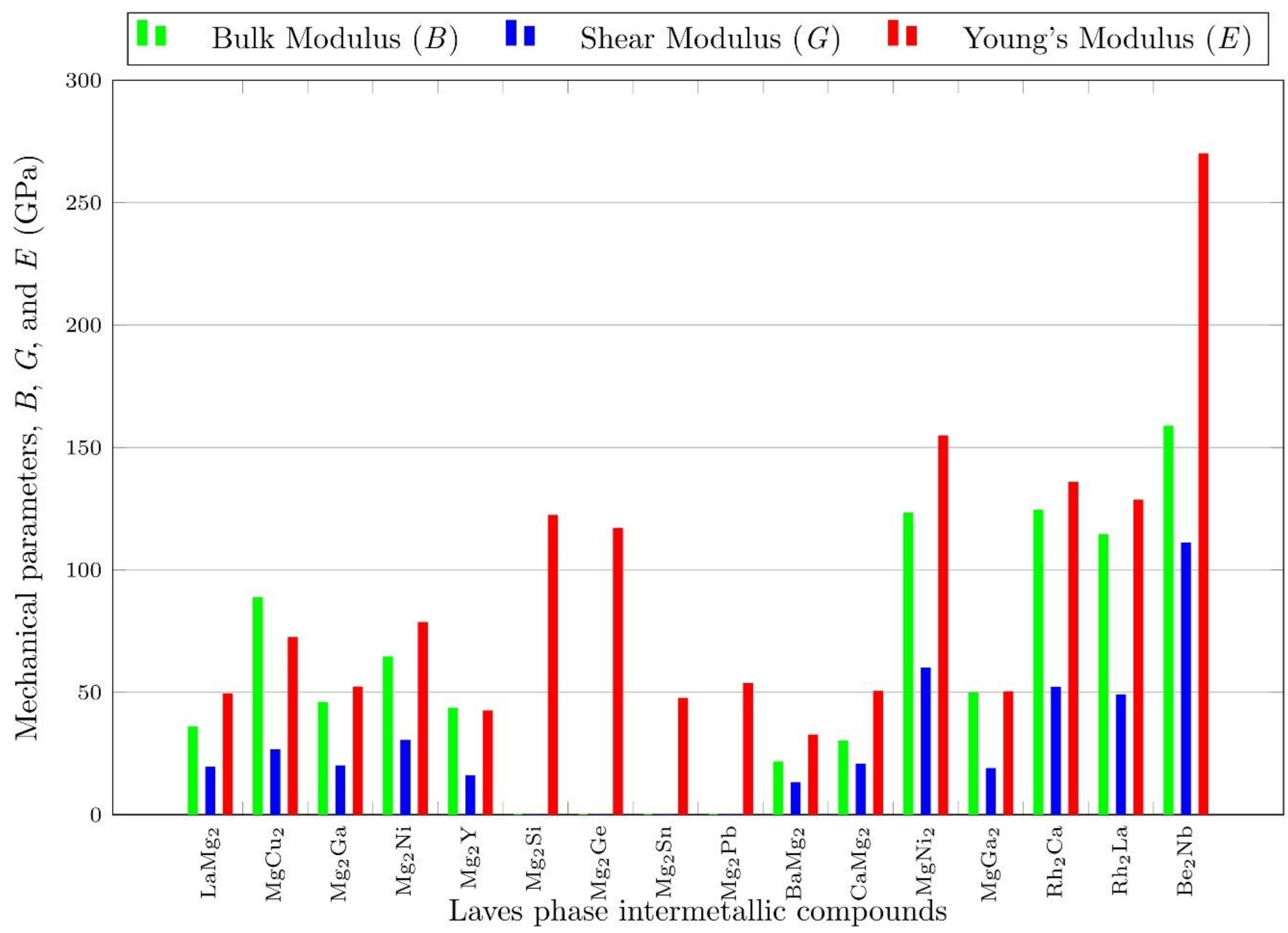

**Figure 5:** Best performing Laves phase intermetallics in terms of mechanical properties.

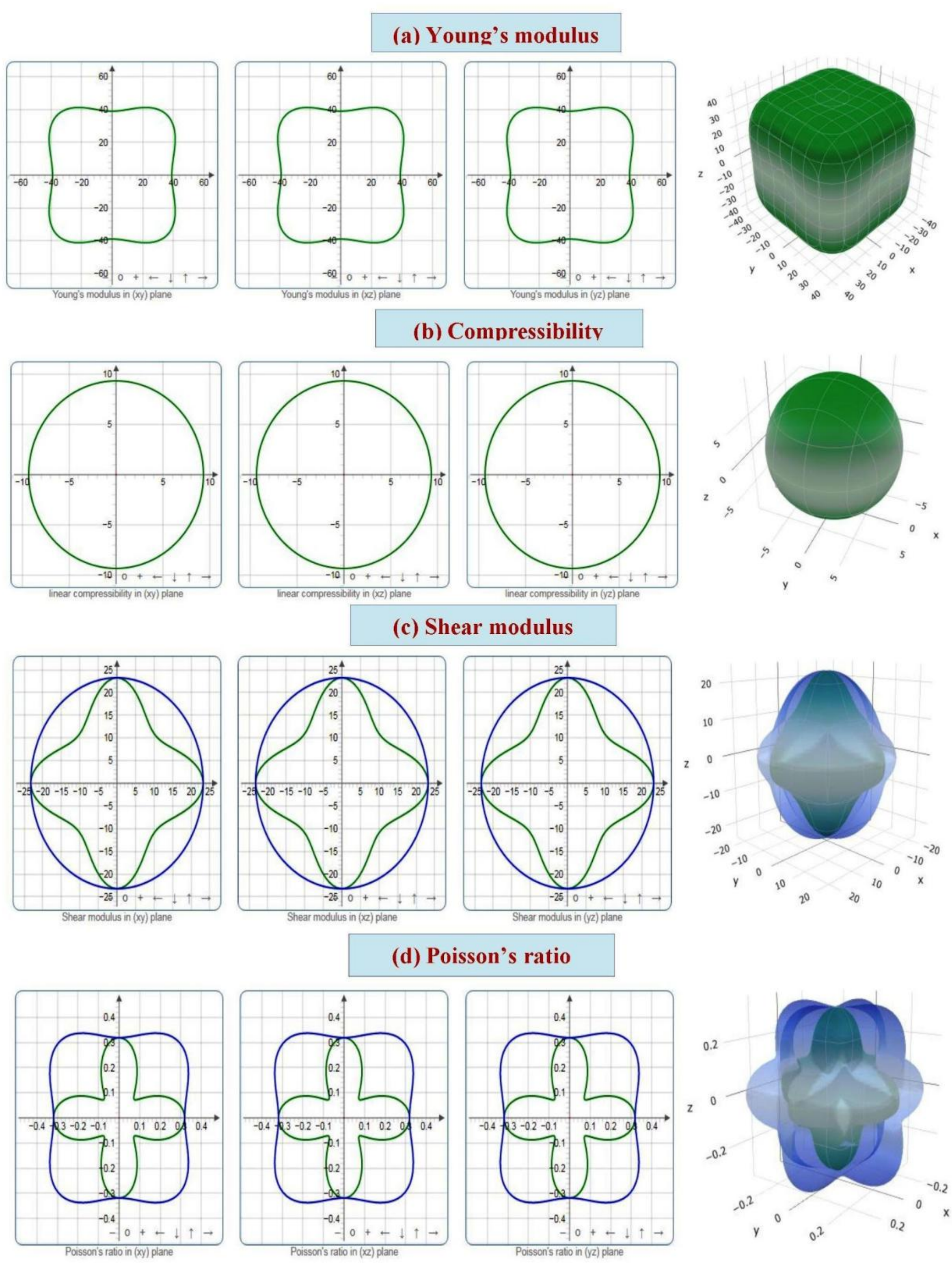

**Figure 6:** The two-dimensional (2D) and three-dimensional (3D) plots of (a) Young's modulus, (b) linear compressibility, (c) shear modulus, and (d) Poisson's ratio of $LaMg_2$ Laves phase.

**Tables**

**Table 1:** The calculated mechanical properties of rare-earth Laves phase $LaMg_2$ in comparison with other members in this intermetallic family.

| ***Compounds*** | $C_{11}$ | $C_{12}$ | $C_{44}$ | $C_{12}$-$C_{44}$ | ***B* (GPa)** | ***G* (GPa)** | ***E* (GPa)** | ***B*/*G*** | *v* | ***Ref.*** |
|---|---|---|---|---|---|---|---|---|---|---|
| $LaMg_2$ | 55.50 | 25.94 | 23.24 | 2.70 | 35.80 | 19.38 | 49.26 | 1.84 | 0.27 | This work |
| | 58.00 | 25.00 | 22.20 | - | 36.00 | 19.08 | 49.09 | - | - | [16] |
| | 58.04 | 24.09 | 21.08 | - | 36.00 | 19.07 | 50.01 | - | - | [22] |
| $MgCu_2$ | 107.9 | 79.0 | 34.6 | - | 88.6 | 26.5 | 72.3 | 3.34 | - | [22] |
| $Mg_2Ga$ | - | - | - | - | 45.8 | 19.9 | 52.1 | 2.31 | - | [22] |
| $Mg_2Ni$ | - | - | - | - | 64.4 | 30.3 | 78.5 | 2.13 | - | [23] |
| $Mg_2Y$ | - | - | - | - | 43.46 | 15.83 | 42.36 | - | 0.33 | [23] |
| $Mg_2Si$ | 121.26 | 27.12 | 47.43 | - | - | - | 122.17 | - | - | [24] |
| $Mg_2Ge$ | 113.71 | 24.14 | 45.05 | - | - | - | 116.92 | - | - | [24] |
| $Mg_2Sn$ | 68.56 | 28.72 | 14.88 | - | - | - | 47.47 | - | - | [24] |
| $Mg_2Pb$ | 62.44 | 30.28 | 21.22 | - | - | - | 53.50 | - | - | [24] |
| $BaMg_2$ | - | - | - | - | 21.5 | 13.0 | 32.5 | 1.65 | - | [22] |
| $CaMg_2$ | - | - | - | - | 30.1 | 20.6 | 50.3 | 1.46 | - | [22] |
| $MgNi_2$ | - | - | - | - | 123.2 | 59.9 | 154.7 | 2.05 | - | [22] |
| $MgGa_2$ | - | - | - | - | 49.8 | 18.8 | 50.2 | 2.64 | - | [22] |
| $Rh_2Ca$ | 175 | 99 | 64 | 35 | 124.33 | 51.93 | 135.75 | 2.39 | 0.31 | [9] |
| $Rh_2La$ | 157 | 93 | 65 | 28 | 114.33 | 48.91 | 128.42 | 2.34 | 0.31 | [9] |
| $Be_2Nb$ | 309.50 | 83.22 | 109.46 | - | 158.65 | 110.93 | 269.88 | 1.43 | 0.21 | [7] |

**Table 2:** The minimum and maximum values of the Young's modulus, Linear compressibility, Shear modulus, Poisson's ratio, and the ratio A of $LaMg_2$.

| $Y_{min}$ (GPa) | $Y_{max}$ (GPa) | $A_Y$ | $K_{min}$ ($TPa^{-1}$) | $K_{max}$ ($TPa^{-1}$) | $A_K$ | $G_{min}$ (GPa) | $G_{max}$ (GPa) | $A_G$ | $\nu_{min}$ | $\nu_{max}$ | $A_\nu$ |
|---|---|---|---|---|---|---|---|---|---|---|---|
| 38.98 | 57.31 | 1.47 | 9.31 | 9.31 | 1.00 | 14.78 | 23.24 | 1.57 | 0.103 | 0.419 | 4.05 |